\documentclass[pre,twocolumn,aps,showpacs,amsmath,amsfonts,
eqsecnum]{revtex4}
\usepackage{graphicx}
\usepackage{bm}
\newcommand{\Onecol} {\begin{widetext} \onecolumngrid} 
\newcommand{\Twocol} {\end{widetext} \twocolumngrid} 

 \renewcommand{\it}[1]{\textit{#1}}

\newcommand{\BE}[1]{\begin{equation}\label{eq:#1}} 
\newcommand{\EE}{\end{equation}}
\newcommand{\BEA}[1]{\begin{eqnarray} \label{eq:#1}} 
\newcommand{\EEA}{\end{eqnarray}} 
\newcommand{\BEa}{\begin{eqnarray*}} 
\newcommand{\EEa}{\end{eqnarray*}} 
\def\nn {\nonumber} \def\br {\\ \nonumber} 
\newcommand{\B}[1]{{\bm{#1}}}
\def\<{\left \langle} \def\>{\right\rangle}
\newcommand{\ok}{(\omega,\B k)}
\newcommand{\tr}{(t,\B r)}

\newcommand{\Sb}[1]{_{_{\text {#1}}}} 
\newcommand{\Sp}[1]{^{^{\text {#1}}}} 
\def\Fbox#1{\vskip1ex\hbox to 8.5cm{\hfil\fboxsep0.3cm\fbox{%
\parbox{8.0cm}{#1}}\hfil}\vskip1ex\noindent} 
\def\K41{\fboxrule0.2ex\fbox{\large\text{K41}}}
\def\Fbox#1{\vskip1ex\hbox to 8.5cm{\hfil\fboxsep0.3cm\fbox{%
\parbox{8.0cm}{#1}}\hfil}\vskip1ex\noindent} 
\begin{document}
\title{Clustering of inertial particles in a turbulent flow}
\author{Tov Elperin$^1$}
\email{elperin@menix.bgu.ac.il}
\homepage{http://www.bgu.ac.il/~elperin}
\author{Nathan Kleeorin$^1$}
\email{nat@menix.bgu.ac.il}
\author{Michael~A.~Liberman$^2$}
\email{Michael.Liberman@dec11.fysik.uu.se}
\homepage{http://lili.fysik.uu.se/}
\author{Victor S. L'vov$^3$}
\email{Victor.Lvov@Weizmann.ac.il}
\homepage{http://lvov.weizmann.ac.il}
\author{Anna Pomyalov$^3$}
\email{Anna.Pomyalov@Weizmann.ac.il}
\author{Igor Rogachevskii$^1$}
\email{gary@menix.bgu.ac.il} \homepage{http://www.bgu.ac.il/~gary}
\affiliation{$^1$The Pearlstone Center for Aeronautical
Engineering
Studies, Department of Mechanical Engineering,\\
Ben-Gurion University of the Negev, P. O. Box 653, Beer-Sheva
84105, Israel\\
$^2$Department of Physics, Uppsala University, Box 530, SE-751
21, Uppsala, Sweden\\
$^3$Department of Chemical Physics, The Weizmann Institute of
Science, Rehovot 76100, Israel}
\begin{abstract}
We analyzed formation of small-scale inhomogeneities of particle
spatial distribution (particle clustering) in a turbulent flow.
The particle clustering is a consequence of a spontaneous
breakdown of their homogeneous space distribution, and is caused
by a combined effect of the particle inertia and a finite
correlation time of the turbulent velocity field. Theory of the
particle clustering is extended to the case when the particle
Stokes time is larger than the Kolmogorov time scale, but is much
smaller than the correlation time at the integral scale of
turbulence. The criterion of the clustering instability is
obtained. Applications of the analyzed effects to the dynamics of
inertial particles in industrial turbulent flows are discussed.
\end{abstract}
\date{\today}
\maketitle

\section{Introduction}

It is generally believed that turbulence promotes mixing (see,
e.g., \cite{10,C80,PS83,McC90,S96,BL97,W00,S01,BH03}). However,
laboratory experiments and observations in the atmospheric
turbulence show formation of long-living inhomogeneities in
concentration distribution of small inertial particles and
droplets in turbulent fluid flows (see, e.g., \cite{5,7,4}). The
origin of these inhomogeneities is not always clear but there
influence on the mixing can be hardly overestimated.

The goal of this study is to analyze the particle-air interaction
leading to the formation of strong inhomogeneities of particle
distribution, referred to as \emph{particle clustering}(see
\cite{96-PRL-EKR,02-Our-Clust-PRE}, and references therein).
Particle clustering is a consequence of a spontaneous breakdown of
their homogeneous space distribution. As a result of the nonlinear
stage of clustering, the local density of particles may rise by
orders of magnitude and strongly increase the probability of
particle-particle collisions.

It was suggested in \cite{96-PRL-EKR,14,14A} that the main reason
for the particle clustering is their inertia: the particles inside
the turbulent eddies are carried out to the boundary regions
between the eddies by the inertial forces. This mechanism of the
preferential concentration acts in all scales of turbulence,
increasing toward small scales. Later, this was contested in
Refs.~\cite{15,19} using the so-called "Kraichnan model" \cite{21}
of turbulent advection by the delta-correlated in time random
velocity field, whereby the clustering instability did not occur.
However, it was shown in Ref.~\cite{02-Our-Clust-PRE} that
accounting for a finiteness of correlation time of the fluid
velocity field results in the clustering instability of heavy
particles. The theory of the clustering instability of the
inertial particles advected by a turbulent velocity field was
developed in Ref.~\cite{02-Our-Clust-PRE} and applied to the
dynamics of aerosols in the turbulent atmosphere.

In the present study we extended the theory of particle clustering
to the case when the particle Stokes time is larger than the
Kolmogorov time scale, but is much smaller than the correlation
time at the integral scale of turbulence. The paper is organized
as follows. In Sec.~\ref{s:clust} we present a qualitative
analysis of the clustering instability that causes formation of
particle clusters in a turbulent flow. In Sec.~\ref{s:crit-val} we
evaluate the characteristic parameters that affect clusters
formation as a result of the clustering instability: the particle
response time $\tau_p$, the Kolmogorov micro-scale time
$\tau_\eta$, the characteristic velocity of small and large
particles in turbulent fluid; the degree of compressibility of the
particle velocity field, $\sigma_v(a)$, where $a$ is the particle
size. In Sec.~\ref{ss:particle-vel} we study the velocity of small
and large particles in the turbulent fluid, required for
evaluation of the effect of turbulent diffusion in these two
regimes. In Sec.~\ref{moments} we present a quantitative analysis
for the clustering instability of the second moment of particle
number density. This allows us to generalize the criterion of the
clustering instability, obtained in Ref.~\cite{02-Our-Clust-PRE}.
Finally, in Sec.~\ref{s:disc} we overview the nonlinear effects
which lead to saturation of the clustering instability and
determine the particle number density in the cluster. In
Sec.~\ref{s:disc} we also present numerical estimates for droplet
dynamics for the conditions pertinent to turbulence in diesel
engines.

\section{Clustering of particles in turbulent gas}
\label{s:clust}

\subsection{Basic Equations}
To analyze dynamics of particles we  use the standard continuous
media approximation, introducing the number density $n(t,
\mathbf{r}) $ of spherical particles with radius $a$.

The particles are advected by a turbulent velocity field
${\mathbf{u}}(t,\mathbf{r}) $. Since typically the velocity of the
carrier gas is much smaller than the sound velocity, the
incompressibility constrain $\text{div}\,
{\mathbf{u}}(t,{\mathbf{r}}) = 0 $ is applicable. The particle
material density $\rho_p$ is much larger than the density $\rho$
of the ambient fluid. For heavy particles
${\mathbf{v}}(t,{\mathbf{r}})\neq{\mathbf{u}}(t,{\mathbf{r}}) $
due to the particle inertia \cite{MR83,M87,M96}. Therefore, the
compressibility of the particle velocity field
${\mathbf{v}}(t,{\mathbf{r}}) $ must be taken into account (see
\cite{96-PRL-EKR,14,14A,02-Our-Clust-PRE}), since the growth rate
of the clustering instability, $\gamma $, is proportional to
$\langle | \text{div} \, {\mathbf{v}}(t,{\mathbf{r}})|^{2}\rangle
$, where $\langle \cdot \rangle $ denotes ensemble average.

Let $\Theta(t,{\mathbf{r}}) $ be the deviation of the particle
number density $n(t, \mathbf{r}) $ from its uniform mean value
$\bar{n} \equiv \langle n \rangle $:
\begin{equation}
\Theta(t,{\mathbf{r}})=n(t, \mathbf{r})-\bar{n}\,. \label{A2}
\end{equation}
The pair correlation function of $\Theta(t,{\mathbf{r}}) $ is defined
as
\begin{equation}
\Phi({\mathbf{R}},{\mathbf{r}},t)\equiv\langle
\Theta(t,{\mathbf{r+R}})\Theta(t,{\mathbf{r}})\rangle\ . \label{2}
\end{equation}
For the sake of simplicity we will consider only a spatially
homogeneous, isotropic case when
$\Phi({\mathbf{R}},{\mathbf{r}},t) $ depends only on the
separation distance $R$ and time $t$:
\begin{equation}
\Phi(t,{\mathbf{R}},{\mathbf{r}})\rightarrow\Phi(t,R)\ .
\label{02-Our-Clsut-PRE}
\end{equation}
Denote the probability of the pair collisions between particles as
$p(t) $ which can be expressed as
\begin{equation}
p(t)\propto c \left\{1+\frac{\langle[n(t,
\mathbf{r})]^{2}\rangle}{\bar{n}^{2}} \right\} = c
\left[1+\frac{\Phi(t,0)}{\bar{n}^{2}} \right]\ . \label{4}
\end{equation}
Obviously, a large increase of $\Phi(t,R) $ above the level of
$\bar{n}^{2} $ leads to a strong grows in the frequency of the
particle collisions.

In the analytical treatment of the problem we use the standard
equation for $n(t, \mathbf{r}) $:
\begin{equation}
\frac{\partial n(t, \mathbf{r})}{\partial
t}+\mathbf{\nabla}\cdot[n(t,
\mathbf{r}){\mathbf{v}}(t,{\mathbf{r}})]=D\triangle n(t,
\mathbf{r}), \label{5}
\end{equation}
where D is the coefficient of molecular (Brownian) diffusion. The
equation for $\Theta(t,\mathbf{r}) $ follows from Eq.~(\ref{5}):
\begin{eqnarray}\label{6}
\frac{\partial \Theta(t, \mathbf{r})}{\partial
t}&+&[{\mathbf{v}}(t,{\mathbf{r}})\cdot\nabla]\Theta(t,
\mathbf{r})\\ \nonumber &=&-\Theta(t, \mathbf{r}) \, \text{div} \,
{\mathbf{v}}(t,{\mathbf{r}})+D\triangle \Theta(t, \mathbf{r})\ .
\end{eqnarray}
Here we neglected the term $\propto\bar{n} \,
\text{div}\,{\mathbf{v}} $, describing the effect of an external
source of fluctuations. It was shown in
Ref.~\cite{02-Our-Clust-PRE} that this effect is usually much
smaller than the effect of self-excitation of fluctuations of
particle number density.

One can use Eq.~(\ref{6}) to derive equation for $\Phi(t,R) $ by
averaging the equation for
$\Theta(t,{\mathbf{r+R}})\Theta(t,{\mathbf{r}}) $ over statistics
of the advected turbulent velocity field
${\mathbf{v}}(t,{\mathbf{r}}) $. In general this procedure is
quite involved even for simple models of the advecting velocity
fields, see, e.g., Ref.~\cite{02-Our-Clust-PRE}. Nevertheless, the
qualitative understanding of the underlying physics of the
clustering instability, leading to both, the exponential growth of
$\Phi(t,R) $ and its nonlinear saturation, can be elucidated by a
more simple and transparent analysis, that is presented below.

\subsection{Qualitative analysis of the clustering instability}
\subsubsection{On Richardson-Kolmogorov cascade theory of
turbulence}

In our discussion we will use the well known Richardson-Kolmogorov
cascade theory of turbulence (see, e.g., Refs.~\cite{9,10,11}).
For the large Reynolds numbers ${\cal R}\!e \gg 1 $ the
characteristic scale $L $ of energy injection (outer scale) is
much larger than the length of the dissipation scales
(\emph{viscous scale} $\eta $) $L\gg\eta $. In the so-called
\emph{inertial interval} of scales, where $L>r>\eta $, the
statistics of turbulence within the Kolmogorov theory is governed
by the only dimensional parameter, $\varepsilon $, the rate of the
turbulent energy dissipation. Then, the velocity $u(r) $ and the
energy of turbulent motion $E(r) $ at the characteristic scale $r
$ (referred below as r-eddies) may be found by the dimensional
reasoning:
\begin{equation}
u(r)\approx(\varepsilon r)^{1/3}, \quad
E(r)=\frac{1}{2}\rho[u(r)]^{2}. \label{7}
\end{equation}
Similarly, the turnover time of $r $-eddies, $\tau(r) $, which is
of the order of their life time, may be estimated as
\begin{equation}
\tau(r)\approx r/u(r)\approx \varepsilon^{-1/3}r^{2/3}.
\label{96-PRL-EKR}
\end{equation}
To elucidate the clustering instability let us consider a cluster
of particles with a characteristic scale $\ell $ moving with the
velocity ${\mathbf{V}}_\text{cl}(t) $. The scale $\ell $ is a
parameter which governs the growth rate of the clustering
instability, $\gamma $. It sets the bounds for two distinct
intervals of scales: $L>r>\ell $ and $\ell>r>\eta $. However, if
the size of particles $a$ is larger than the viscous scale $\eta$,
the second range of scales becomes $\ell>r>a $, because we cannot
consider scales which are smaller than the size of particles.
Large $r $-eddies with $r>\ell $ sweep the $\ell $-cluster as a
whole and determine the value of ${\mathbf{V}}_\text{cl}(t) $.
This results in the diffusion of the clusters, and eventually
affects their distribution in a turbulent flow. The small $r
$-eddies determine the dynamics of particles inside the cluster.
The role of these eddies is multifold. First, they lead to the
turbulent diffusion of the particles within the scale of a cluster
size. Second, due to the particle inertia they tend to accumulate
particles in the regions with small vorticity, which leads to the
preferential concentration of the particles. Third, the particle
inertia also causes a transport of fluctuations of particle number
density from smaller scales to larger scales, i.e., in regions
with larger turbulent diffusion. The latter can decrease the
growth rate of the clustering instability (see below). Thus, the
clustering is determined by the competition between these three
processes.

\subsubsection{Effect of turbulent diffusion}

We consider Eq.~(\ref{6}) for $\Theta(t,\mathbf{r}) $ in the
coordinate system co-moving with the $\ell $-cluster and assume
${\mathbf{v}}(t,{\mathbf{r}}) \approx {\mathbf{u}}(t,{\mathbf{r}})
$. In this reference frame the advectived velocity
${\mathbf{v}}(t,{\mathbf{r}}) $ should be replaced by
$[{\mathbf{v}}(t,{\mathbf{r}})-{\mathbf{V}}_\text{cl}(t)] $. In
particular, the advection term in Eq.~(\ref{6}) takes the form
\begin{equation}
\text{Adv} \equiv \left\{[{\mathbf{v}}(t,{\mathbf{r}}) -
{\mathbf{V}}_\text{cl}(t)]\cdot\nabla\right\}
\Theta(t,\mathbf{r})\ . \label{9}
\end{equation}
Averaging this term over statistics of turbulent velocity field
leads to the turbulent diffusion of particles. It is well known
(see, e.g. Ref.~\cite{10}) that the turbulent diffusion can be
modelled by the renormalization of the diffusion term in the right
hand side (RHS) of Eq.~(\ref{6}):
\begin{equation}
D\Rightarrow D+D_{_\text{T}}, \label{10}
\end{equation}
where $D_{_\text{T}} $ is the turbulent diffusion coefficient. The
main contribution to the advected velocity in Eq.~(\ref{9}) is due
to the velocity of $\ell $-eddies, $v(\ell) $. Therefore
$D_{_\text{T}} $ becomes a function of scale $\ell $,
$D_{_\text{T}}\Rightarrow D_{_\text{T}}(\ell) $, and can be
estimated from the parameters of $\ell $-eddies, using dimensional
reasoning:
\begin{equation}
D_{_\text{T}}(\ell)\approx\frac{1}{3}\ell v(\ell)\ . \label{11}
\end{equation}
Here we used the commonly accepted \cite{9,11} numerical factor
$\frac{1}{3} $ for the turbulent diffusion coefficient in an
isotropic turbulence. Now the part of Eq.~(\ref{6}) describing the
turbulent diffusion may be written as
\begin{equation}
\frac{\partial \Theta(t, \mathbf{r})}{\partial
t}=-D_{_\text{T}}(\ell)\triangle\Theta(t, \mathbf{r})\ .
\label{00Bel}
\end{equation}

During the linear stage of the cluster evolution the particle
distribution inside the cluster does not change. Therefore, the
function $\Theta(t, \mathbf{r}) $ in the expression for the
correlation function $\Phi$ can be factorized:
\begin{equation}
\Theta(t, \mathbf{r}) \Rightarrow A_{\ell}(t)\theta_{\ell}(r)\ .
\label{13}
\end{equation}
Then, Eq.~(\ref{00Bel}) yields
\begin{equation}
A_{\ell}^{2}(t) = A_{\ell}^{2}(0)\exp[-\gamma_\text{dif} \, t]\, ,
\ \gamma_\text{dif}= \frac{2D_{_\text{T}}(\ell)}
{\ell^{2}}\approx\frac{2v(\ell)}{3\ell}\,, \label{14}
\end{equation}
where the Laplace operator is evaluated as $1/\ell^{2} $ and we
used Eq.~(\ref{11}).

\subsubsection{Effect of particles inertia}

For the qualitative analysis of the particle inertia we consider
Eq.~(\ref{6}), written in the co-moving reference frame, taking
into account only inertia term:
\begin{equation}
\frac{\partial \Theta(t, \mathbf{r})}{\partial t}=-\Theta(t,
\mathbf{r}) \, \text{div}[{\mathbf{v}}(t,{\mathbf{r}})
-{\mathbf{V}}_\text{cl}(t)]\ . \label{15}
\end{equation}
As before we can factorize the function $\Theta(t, \mathbf{r}) $
according to Eq.~(\ref{13}) and simplify the partial differential
equation ~(\ref{15}), reducing it to the differential equation for the
cluster amplitude $A_\ell(t) $:
\begin{equation}
\frac{dA_{\ell}(t)}{dt}=-A_{\ell}(t)\, b_{\ell}(t)\ . \label{16}
\end{equation}
Here we neglected the $\mathbf{r} $-dependence of the divergence
term inside the cluster in the RHS of Eq.~(\ref{15}), so that this
term becomes a function of $t $ only:
\begin{equation}
\text{div}[{\mathbf{v}}(t,{\mathbf{r}})-{\mathbf{V}}_\text{cl}(t)]
\Rightarrow b_{\ell}(t)\ .\label{17}
\end{equation}

In the turbulent flow field the function $b_{\ell}(t) $ may be
considered as a random process with some correlation time
$\tau_{b} $, which will be evaluated below. Since the instability
of $\ell $-clusters is caused by $\ell $-eddies, the correlation
time $\tau_{b} $ can be estimated as the turnover time of the
particle velocity field of $\ell$-eddies:
\begin{equation}
\tau_{b}\approx \ell/v(\ell)\ . \label{18}
\end{equation}

The evaluation of the mean square value of $b_{\ell}(t) $ requires
a more careful consideration. One cannot estimate $\text{div} \,
\mathbf{v}(\mathbf{r}) $ by the dimensional reasoning as $v(r)/r
$. Indeed, in the incompressible flow $\text{div}\,
{\mathbf{v}(\mathbf{r})}=0 $. To elucidate this issue we introduce
a dimensionless parameter $\sigma_v $, a \emph{degree of
compressibility} of the velocity field of particles in $\ell
$-clusters, ${\mathbf{v}}_{\ell}(t,{\mathbf{r}}) $, defined by
\begin{equation}
\sigma_v\equiv\langle[\text{div} \,
{\mathbf{v}}_{\ell}]^{2}\rangle/ \langle|\nabla
\times{\mathbf{v}}_{\ell}|^{2}\rangle \ . \label{19}
\end{equation}
This parameter may be of the order of 1 (see Refs.~\cite{14,15}).
At the moment we can evaluate $b_{\ell}(t) $ via this yet unknown
parameter $\sigma_v $ as follows:
\begin{equation}
\langle b_{\ell}(t)\rangle=0, \quad \langle
b_{\ell}^{2}(t)\rangle\approx\sigma_v[v(\ell)/\ell]^{2}\ .
\label{20}
\end{equation}
Let us show that the stochastic differential equation ~(\ref{16})
results in an exponential growth of $\langle
A_{\ell}^{2}(t)\rangle $, i.e. in the instability. Indeed, the
solution of this equation reads:
\begin{equation}
A_{\ell}(t) = A_{\ell}(0)\exp[-I(t)]\,, \quad
I(t)\equiv\mathrel{\mathop{\stackrel{t}{\int}}\limits_{0}}
b_{\ell}(\tau)d\tau\ . \label{21}
\end{equation}
Integral $I(t) $ may be written as the sum of integrals $I_{n} $
over small time intervals $\tau_{b} $:
\begin{equation}
I(t)=\mathrel{\mathop{\stackrel{N}{\sum}}\limits_{n=1}}I_{n}\,,
\quad I_{n} \equiv\mathrel{\mathop{\stackrel{(n+1)\tau_{b}}{\int}}
\limits_{n\tau_{b}}}b_{\ell}(\tau)d\tau, \quad
N=\frac{t}{\tau_{b}}\ . \label{22}
\end{equation}
By definition, $\tau_{b} $ is the correlation time of the random
process $b_{\ell}(t) $. Therefore the integrals $I_{n} $ can be
considered as independent random variables. According to the
central limit theorem, the sum of a large number of statistically
independent random variables is distributed as the Gaussian random
variable. Therefore the total integral $I(t) $ can be estimated as
\begin{equation}
I(t)\simeq\sqrt{N\langle I_{n}^{2}\rangle}\, \tilde \zeta\, ,
\quad \langle I_{n}^{2}\rangle=\tau_{b}^{2}\langle
b_{\ell}^{2}(t)\rangle\ . \label{23}
\end{equation}
Here $\tilde \zeta $ is a Gaussian random variable with zero mean
and unit variance. The probability density function of $\tilde
\zeta $ reads
\begin{equation}
{\mathcal{P}}(\tilde
\zeta)=\frac{1}{\sqrt{2\pi}}\exp\left(-\frac{\tilde \zeta^{2}}{2}
\right ). \label{24}
\end{equation}
Using Eqs. (\ref{21}), (\ref{24}) and (\ref{22}) we obtain
\begin{eqnarray}\nonumber
\langle A_{\ell}^{2}(t)\rangle & \simeq &
A_{\ell}^{2}(0)\int\exp[-2\sqrt{N\langle I_{n}^{2}\rangle}\tilde
\zeta]{\mathcal{P}}(\tilde \zeta)d\tilde \zeta \\
& = & A_{\ell}^{2}(0)\exp[\gamma_{in} t]\, , \label{25}
\end{eqnarray}
where the growth rate $\gamma_{in} $ is given by
\begin{equation}
\gamma_\text{in}=\frac{N\langle I_{n}^{2}\rangle}{\tau_{b}}\approx
2\tau_{b}\langle b_{\ell}^{2}(t)\rangle\approx \frac{2\sigma_v\, v
(\ell)}{\ell}\ . \label{26}
\end{equation}
In the last estimate we used Eqs. (\ref{18}) and (\ref{20}). It is
clearly seen that the source of the instability is a nonzero value
of $\sigma_v $, i.e. a \emph{compressibility of the particle
velocity field}, ${\mathbf{v}}(t,{\mathbf{r}}) $.

In the case of the small enough particles, $\tau_p\leq\tau(\eta)
$, where $\tau(\eta) $ is the turnover time of the smallest,
Kolmogorov micro-scale eddies. In this case all the particles are
almost fully involved in turbulent motion, and one concludes that
$u(t,{\mathbf{r}})\approx v(t,{\mathbf{r}}) $ and $v(\ell)\approx
u(\ell) $. Also we will see below that in this case the largest
value of $\sigma_v $ is attained for $\ell \approx\eta $. Hence
for $\tau_p\leq\tau(\eta) $ the most unstable clusters are those
with $\ell \approx\eta $. In summary:

\textbullet~In the case $\tau_p\le \tau (\eta)$, the
characteristic scale of the most unstable clusters of small enough
particles is of the order of Kolmogorov micro-scale of turbulence,
$\eta $.

\textbullet~The characteristic growth rate of the clustering
instability is of the order of the turnover frequency of $\eta
$-eddies, $1/\tau(\eta) $.

\textbullet~The particle clusters are unstable for heavy enough
particles, such that the degree of compressibility $\sigma_v(\eta)
$ of their effective advective velocity field exceeds some
threshold value $\sigma_\text{cr}\approx 0.3 $ (see
Ref.~\cite{02-Our-Clust-PRE}).

\section{The particle velocity field}
\label{s:crit-val}

In this section we discuss the compressibility of the particle
velocity field. In particular, to clarify the range of validity of
the Stokes approximation for the particle motion in fluid, in
Sec.~\ref{ss:times} we evaluate the particle response time
$\tau_p$ as compared with the Kolmogorov micro-scale time
$\tau_\eta$.  In Sec.~\ref{ss:particle-vel} we study the velocity
of small and large particles in the turbulent fluid, required for
evaluation of the effect of turbulent diffusion in these two
regimes. This study also allows us to find the dependence of the
degree of compressibility of the particle velocity field,
$\sigma_v$, on the particle response time.

\subsection{Characteristic time scales and validly
of the Stokes approximation} \label{ss:times}

Let us assume that the radius of the particles is small, so that
the particle Reynolds number ${\mathcal{R}}e_p $ is smaller than
the critical value, ${\mathcal{R}}e_\text{cr}$,  at which the
laminar flow over a particle looses its stability. Then, we can
apply the Stokes approximation. It states that  the fluid-particle
friction force is proportional to the\emph{ slip velocity}, the
difference between the particle velocity and the fluid velocity. A
careful analysis performed by Lumley \cite{6} shows that in a
turbulent flow the validity condition for
${\mathcal{R}}e_p\leq{\mathcal{R}}e_\text{cr} $ may be expressed
via the particle radius $a $ and the Kolmogorov micro-scale $\eta
$ as follows:
\begin{equation}
a\leq2\eta(\rho/\rho_p)^{1/3}. \label{30}
\end{equation}
In the present analysis the ratio of the inertial time scale of
the particles (the Stokes time scale $\tau_p $) and the turnover
time of $\eta $-eddies in the Kolmogorov micro-scale $\tau(\eta)
,$ is of primary importance. The particle response time is given
by
\begin{equation}
\tau_p=\frac{m_p}{6\pi\nu\rho a}=\frac{2\rho_p a^{2}}{9\rho\nu}\,,
\label{31}
\end{equation}
where the particle mass $m_p $ is:
\begin{equation}
m_p=\frac{4\pi}{3}a^{3}\rho_p\ . \label{32}
\end{equation}
The Kolmogorov micro-scale $\eta $ is determined from the
condition that the Reynolds number for eddies of scale $\eta $ is
equal to 1:
\begin{equation}
{\mathcal{R}}e_{\eta}=\eta v(\eta)/\nu=1\ . \label{33}
\end{equation}
Here $v(\eta) $ is the characteristic velocity of $\eta $-scale
eddies. It is related to the turnover time of the eddies as
$\tau(\eta)=\eta/v(\eta) $. This allows us to rewrite the
expression (\ref{33}) as follows:
\begin{equation}
\tau(\eta)=\eta^{2}/\nu. \label{34}
\end{equation}
Then, the ratio of the time-scales $\tau_p $ and $\tau(\eta) $
follows from Eqs. (\ref{31}) and (\ref{34}):
\begin{equation}
\frac{\tau_p}{\tau(\eta)}=\frac{2\rho_p a^{2}} {9\rho\eta^{2}}\ .
\label{35}
\end{equation}
Substituting the condition (\ref{30}) for the validity of the
Stokes approximation we find
\begin{equation}
\frac{\tau_p}{\tau(\eta)}\leq\left(\frac{\rho_p}
{\rho}\right)^{1/3}\,, \label{36}
\end{equation}
Equation (\ref{36}) implies that for "heavy" particles in a gas,
that satisfy the Stokes approximation, the particle response time
scale may be about ten times larger than the Kolmogorov time
scale: $\tau_p\leq 10\tau(\eta) $.

\subsection{Particle velocity in turbulent fluid}
\label{ss:particle-vel}
\subsubsection{Equation for a particle velocity}

Assuming that the particles are small enough such that
Eq.~(\ref{30}) is valid, we can apply Stokes' law for the
fluid-particle friction force ${\mathbf{F}}_p(t,{\mathbf{r}}) $:
\begin{equation}
{\mathbf{F}}_p(t,{\mathbf{r}})=\zeta[{\mathbf{v}}_p
(t,{\mathbf{r}})-{\mathbf{u}}(t,{\mathbf{r}})]\, , \label{37}
\end{equation}
with the Stokes friction coefficient $\zeta $ given by
\begin{equation}
\zeta=6\,\pi\,\rho\nu\, a\ . \label{38}
\end{equation}
The Newton equation for a particle reads:
\begin{equation}
m_p\frac{d{\mathbf{v}}_p(t,{\mathbf{r}})}{dt}
=-{\mathbf{F}}_p(t,{\mathbf{r}})=\zeta[{\mathbf{u}}(t,
{\mathbf{r}})-{\mathbf{v}}_p(t,{\mathbf{r}})]\ . \label{39}
\end{equation}
Here the total time derivative $(d/dt) $ takes into account the
time dependence of the particle coordinate $\mathbf{r} $:
\begin{equation}
\frac{d}{dt}=\left[\frac{\partial}{\partial
t}+{\mathbf{v}}_p(t,{\mathbf{r}})\cdot\nabla\right]\ . \label{40}
\end{equation}
Now Eq.~(\ref{39}) takes the form:
\begin{equation}
\left\{\tau_p\left[\frac{\partial}{\partial
t}+{\mathbf{v}}_p(t,{\mathbf{r}})\cdot\nabla\right]+1\right\}
{\mathbf{v}}_p(t,{\mathbf{r}})={\mathbf{u}}(t,{\mathbf{r}}) \ .
\label{41}
\end{equation}
In the following we analyze this equation in two limiting cases:
for small particles with $\tau_p $ smaller than the turnover time
of $\ell $-eddies (Sec.~\ref{sss:small}), and for large particles,
in Sec.~\ref{sss:large}.

\subsubsection{Velocity of small particles}
\label{sss:small}

In this section we consider small particles for which the Stokes
time $\tau_p $ is smaller than the turnover time of $\ell
$-eddies, $\tau(\ell)\approx \ell/u(\ell) $. These particles are
completely involved in the motion of $\ell $-eddies and for
$\tau_p/\tau(\ell)=0 $, we can write
${\mathbf{v}}_{\ell}(t,{\mathbf{r}}) =
{\mathbf{u}}_{\ell}(t,{\mathbf{r}}) $. Here subscript "${\ell} $"
denotes that we are dealing with the velocity field of $\ell
$-eddies, ${\mathbf{u}}_{\ell}(t,{\mathbf{r}}) $, and the velocity
of particles in $\ell $-clusters,
${\mathbf{v}}_{\ell}(t,{\mathbf{r}}) $, that was generated by
$\ell $-eddies. In this approximation the velocity field
${\mathbf{v}}_{\ell}(t,{\mathbf{r}}) $ is incompressible.
Therefore, the compressibility parameter $\sigma_v $ may be
determined by the first order corrections.

To find the corrections to ${\mathbf{v}}_{\ell}(t,{\mathbf{r}}) $
linear in $\tau_p $, we consider a formal solution of
Eq.~(\ref{41}) in the form:
\begin{equation}
{\mathbf{v}}_{\ell}(t,{\mathbf{r}})=\left[\tau_p\frac{d}
{dt}+1\right]^{-1}{\mathbf{u}}_{\ell}(t,{\mathbf{r}})\ .
\label{42}
\end{equation}
Here the inverse operator is understood as its Taylor series
expansion:
\begin{equation}
{\mathbf{v}}_{\ell}(t,{\mathbf{r}})={\mathbf{u}}_{\ell}(t,{\mathbf{r}})
-\tau_p\frac{d{\mathbf{u}}_{\ell}(t,{\mathbf{r}})}{dt} +
\frac{\tau_p^{2}}{2} \left[\frac{d{\mathbf{u}}_{\ell}
(t,{\mathbf{r}})}{dt}\right]^{2}+..., \label{43}
\end{equation}
In the linear in $\tau_p $ approximation, this solution becomes
\begin{equation}
{\mathbf{v}}_{\ell}(t,{\mathbf{r}})={\mathbf{u}}_{\ell}(t,
{\mathbf{r}})-\tau_p\left[\frac{\partial}{\partial
t}+{\mathbf{u}}_{\ell}(t,{\mathbf{r}})\cdot\nabla\right]
{\mathbf{u}}_{\ell}(t,{\mathbf{r}})\, , \label{44}
\end{equation}
where in the RHS we replaced ${\mathbf{v}}_{\ell}(t,{\mathbf{r}})
\rightarrow{\mathbf{u}}_{\ell}(t,{\mathbf{r}}) $. Consequently, in
this approximation
\begin{equation}
\text{div}\:{\mathbf{u}}_{\ell}(t,{\mathbf{r}})=
\frac{du_{\ell}^{\alpha}(t,{\mathbf{r}})}
{dr^{\alpha}}\approx\tau_p\frac{du_{\ell}^{\beta}(t,{\mathbf{r}})}
{dr^{\alpha}}\frac{du_{\ell}^{\alpha}(t,{\mathbf{r}})}
{dr^{\beta}}\ \label{45}
\end{equation}
(see Ref.~\cite{96-PRL-EKR}). Equation~(\ref{45}) allows to
determine the compressibility parameter $\sigma_v $, (\ref{19}),
as follows:
\begin{equation}
\sigma_{v} \approx\left[\frac{\tau_p}
{\tau(\ell)}\right]^{2}\approx\left(\frac{2\rho_p}
{9\rho}\right)^{2}\left(\frac{a}{\eta}\right)^{4}
\left(\frac{\eta}{\ell}\right)^{4/3}\, . \label{46}
\end{equation}
In deriving this equation we estimated
$[du_{\ell}^{\alpha}(t,{\mathbf{r}})/dr^{\alpha}] $ as
$[1/\tau({\ell})] $ and used Eqs.~(\ref{35}) and
(\ref{96-PRL-EKR}).

Let $a_{\ast} $ be the characteristic size of particles for which
$\tau_p=\tau(\eta) $ for $\ell=\eta$ and, respectively,
$\sigma_{v}=1 $:
\begin{equation}
a_{\ast}\equiv\eta\sqrt{9\rho/2\rho_p}\ . \label{47}
\end{equation}
Using this notation Eq.~(\ref{46}) can be rewritten as
\begin{eqnarray}
\label{eq:tau-l} \frac{\tau(\ell)}{\tau_p}&\approx&
\left(\frac{a_\ast}a\right)^{2}
\left(\frac{\ell}{\eta} \right)^{2/3} \,, \\
\sigma_{v}&\approx&\left(\frac{a}{a_{\ast}}\right)^{4}
\left(\frac{\eta}{\ell}\right)^{4/3}\ .\label{48}
\end{eqnarray}
Equations (\ref{46}) and (\ref{48}) are valid only if
$\sigma_{v}<1 $, otherwise the approximation of small
$\tau_p/\tau(\ell) $ ratio is violated.

\subsubsection{Velocity of large particles}
\label{sss:large}
\paragraph{Effective equation of motion.}

In this section we consider the opposite case: the large particles
with $\tau_p$ is larger than the turnover time of the smallest
eddies in the Kolmogorov microscale $\tau(\eta)$, but smaller than
the turnover time of the largest eddies $\tau(L)$. Denote by
$\ell_*$ the characteristic scale of eddies for which
\begin{equation}\label{eq:def-l*}
\tau_p= \tau(\ell_*)\ .
\end{equation}
This scale as well as the particles cluster size was introduced in
Ref.~\cite{96-PRL-EKR}. The eddies with $\ell \gg \ell_*$ almost
fully involve particles in their motions, while the eddies with
$\ell \ll \ell_*$ do not affect the particle motions within the
zero order approximation in the ratio $[\tau(\ell)/\tau_p]\ll 1 $.

To determine ${\bm{v}}_{\ell}(t,{\bm{r}}) $ we consider
Eq.~(\ref{41}) in the comoving with $\ell$-eddies frame, where the
surrounding fluid velocity $\B u$ equals to the relative velocity
of the $\ell$-eddy at $\B r$, i.e., $\B u \tr \Rightarrow \B
u_\ell\tr$. At this point one has to take into account that the
$\ell$-eddy is swept out by all $\ell'$-eddies of larger scales, $
\ell'> \ell$, while the particle participates in motions of
$\ell'$-eddies with $ \ell' > \ell_* > \ell$. Therefore, the
relative velocity $ U_{\ell}$ of the $\ell$-eddy and the particle
is determined by $\ell'$-eddies with the intermediate scales,
$\ell_*> \ell' > \ell$. This velocity is determined by the
contribution of $\ell_*$-eddies, and can be considered as a time
and space independent constant $\B u_*$ during the life time of
the $\ell$ eddy and inside it. Velocity $\B u_*$ in our approach
is random and we have to average the final result over statistics
of $\ell_*$-eddies. Then Eq.~(\ref{41}) becomes
\begin{eqnarray}
\left(\tau_p\frac{\partial}{\partial
t}+1\right){\bm{v}}_{\ell}(t,{\bm{r}})
={\bm{u}}_{\ell}(t,{\bm{r}+\B u_* t}) \nonumber\\
- \tau_p\left[{\bm{v}}_{\ell} (t,{\bm{r}})\cdot\nabla\right]
{\bm{v}}_{\ell}(t,{\bm{r}})\ . \label{eq:linn}
\end{eqnarray}
In Eq.~(\ref{eq:linn}) the velocity $\B u_\ell$ is calculated at
point $\B r$ and the velocity $\B v_\ell$ is at $\B r - \B u_*\,
t$. For the sake of convenience we redefine here $\B r - \B u_*\,
t\equiv \B r' \Rightarrow \B r $ and, respectively, $\B r = \B
r'+\B u_*\, t \Rightarrow \B r+\B u_*\, t$.

\paragraph{First non-vanishing contribution to $v_\ell$.}
Clearly,  $ v_\ell \tr \ll u_\ell$ for $\ell \ll \ell_*$.
Therefore we can find the first non-vanishing contribution to
${\bm{v}}_{\ell}(t,{\bm{r}})$ in the limit $[\tau(\ell)\ll
\tau_p]$ by considering the linear version of Eq.~(\ref{eq:linn}):
\begin{equation}
\left(\tau_p\frac{\partial}{\partial
t}+1\right){\bm{v}}_{\ell}(t,{\bm{r}})
={\bm{u}}_{\ell}(t,{\bm{r}}+\B u_* t) \ . \label{eq:v-u-linear}
\end{equation}
In the $\omega, \B k$ representation this equation takes the form:
\begin{equation}
\left(i \omega\, \tau_p +1\right){\bm{v}}_{\ell}\ok
={\bm{u}}_{\ell}(\omega - \B k \cdot \B u_* ,\B k) \,,
\label{eq:v-u-lin-ko}
\end{equation}
that allows one to find the relationship between the second order
correlation functions $F_{v,\ell}^{\alpha\beta}\ok$ and
$F_{u,\ell}^{\alpha\beta}\ok$ of the velocity fields $\B v_\ell$
and $\B u_\ell$:
\begin{equation}\label{eq:Fv-Fu}
F_{v,\ell}^{\alpha\beta}\ok= \frac1{\omega^2 \tau_p^2+1}
F_{u,\ell}^{\alpha\beta}(\omega - \B k \cdot \B u_* ,\B k)\ .
\end{equation}
Functions $F_{u,\ell}^{\alpha\beta}\ok$ and $F_{v,\ell} ^{\alpha
\beta} \ok$ are defined as usual:
\begin{eqnarray}\label{eq:def-Fu}
&& (2 \pi)^4\delta(\omega+\omega')\, \delta (\B k + \B
k')F_{u,\ell}^{\alpha\beta}\ok \\ \nonumber &\equiv& \<
v_\ell^\alpha \ok\, v_\ell^\beta (\omega',\B k')\>\,, \qquad
\text{etc.}
\end{eqnarray}

The simultaneous correlation functions are related to their
$\omega$-dependent counterparts via the integral $\int d \omega/
2\pi$, e.g.,
\begin{equation}\label{eq:rel-1}
F_{v,\ell}^{\alpha\beta}(\B k)=\int \frac{d\omega}{2\pi}
F_{v,\ell}^{\alpha\beta}\ok\ .
\end{equation}
The tensorial structure of $ F_{u,\ell} ^{\alpha\beta}(\B k)$ follows
from the incompressibility condition and the assumption of isotropy:
\begin{equation}\label{eq:incom}
F_{u,\ell} ^{\alpha\beta}(\B k)=P^{\alpha\beta}(\B k)
F_{u,\ell}(k)\,,
\end{equation}
where $P^{\alpha\beta}(\B k)$ is the transversal projector:
\begin{equation}\label{eq:trans}
P^{\alpha\beta}(\B k)=\delta_{\alpha\beta}-k^\alpha k^\beta/k^2\ .
\end{equation}
In the inertial range of scales the function $F_{u,\ell}
^{\alpha\beta}\ok$ may be written in the following form:
\begin{equation}\label{eq:Fu-1}
F_{u,\ell} ^{\alpha\beta}\ok = P^{\alpha\beta}(\B k)\,
F_{u,\ell}(k) \, \tau (\ell)\, f [\omega\, \tau (\ell)]\ .
\end{equation}
Here the dimensionless function $f(x)$ is normalized as follows:
\begin{equation}\label{eq:norm}
\int_{\infty}^\infty f(x)\, dx = 2\pi\ .
\end{equation}

Now we can average Eq.~(\ref{eq:Fv-Fu}) over the statistics of
$\ell_*$-eddies. Denoting the mean value of some function $g(x)$
as $\overline{g(x)}$ we have:
\begin{eqnarray}\label{eq:ev-1}
\overline{f [(\omega- \B k \cdot \B u_*) \,\tau (\ell)]} &\approx&
\overline{\delta [(\omega- \B k \cdot \B u_*)\, \tau(\ell)]} \\
\nonumber &\approx& \frac \ell{\tau(\ell)\, u_*} \, f_*
\left(\frac {u_* \,\omega}\ell\right)\ .
\end{eqnarray}
Here the dimensionless function $f_*(x)$ has one maximum at $x=0$,
and it is normalized according to Eq.~(\ref{eq:norm}). The
particular form of $f_*(x)$ depends on the statistics of
$\ell_*$-eddies and our qualitative analysis is not sensitive to
this form. Thus, we may choose, for instance:
\begin{equation}\label{eq:Lor}
f_*(x)=2 /[x^2+1]\ .
\end{equation}
In Eqs.~(\ref{eq:ev-1}) we took into account that the
characteristic Doppler frequency of $\ell$-eddies (in the random
velocity field $u_*$ of $\ell_*$-eddies) may be evaluated as:
\begin{equation}\label{eq:D-1}
\gamma\Sb D(\ell)\equiv \sqrt{\overline{( \B k \cdot \B u_*)^2 }}
\simeq u_*/\ell\ .
\end{equation}
This frequency is much larger than the characteristic frequency
width of the function $f[\omega \tau(\ell)]$ (equal to
$1/\tau(\ell)$), and therefore the function $f(x)$ in
Eq.~(\ref{eq:ev-1}) may be approximated by the delta function
$\delta(x)$.

After averaging, Eq.~(\ref{eq:Fv-Fu}) may be written as
\begin{equation}\label{eq:Fv}
F_{v,\ell}^{\alpha\beta}\ok= \frac{P^{\alpha\beta}(\B k)
f_*(0)}{\omega^2 \tau_p^2+1}\frac \ell{u_*} F_{u,\ell}(k)\ .
\end{equation}
Here we took into account that $\tau_p\gg \ell/u_*$ that allows us
to neglect the frequency dependence of $ f_*(u_* \,\omega/\ell)$
and to calculate this function at $\omega=0$. Together with
Eq.~(\ref{eq:Fv}) this yields
\begin{equation}\label{eq:rel-2}
F_{v,\ell}^{\alpha\beta}(\B k)=P^{\alpha\beta}(\B k)\,
F_{u,\ell}(k)\frac { \,\ell} {\, \tau_p\, u_*}\,,
\end{equation}
where we used the estimate $ f_*(0)\approx 2$, that follows from
Eq.~(\ref{eq:Lor}).

The equation~(\ref{eq:rel-2}) provides the relationship between
the mean square relative velocity of $\ell$-separated particles,
$v_\ell$, and the velocity of $\ell$-eddies, $u_\ell$:
\begin{eqnarray}\label{eq:v-u-1}
v_\ell\simeq u_\ell \sqrt{\frac { \,\ell} {\, \tau_p\, u_*}}
\simeq u_\ell \sqrt{\frac { \,\ell} {\ell _*}}\ .
\end{eqnarray}

\paragraph{Effective nonlinear equation.}
For a qualitative analysis of the role of the nonlinearity of the
particle behavior in an $\ell $-cluster we evaluate $\nabla $ in
the nonlinear term, Eq.~(\ref{eq:linn}), as $1/\ell $, neglecting
the spatial dependence and the vector structure. The resulting
equation in $\omega $-representation reads:
\begin{eqnarray}\nonumber
(i\omega&+&\gamma_p)V_{\ell}(\omega)=\gamma_
\text{dr}U_{\ell}(\omega)+{\mathcal{N}}_{\omega}\,, \quad
\gamma_p=1/\tau_p\, , \\
{\mathcal{N}}_{\omega}&=&-\frac{1}{2\pi \ell}\int
d\omega_1d\omega_{2}
\delta(\omega+\omega_1+\omega_{2})V_{\ell}(\omega_1)\label{eq:nonlin}
V_{\ell}(\omega_{2})\, ,\nonumber \\
V_{\ell}(\omega)&=&\int v_{\ell}(t)\exp[-i\omega t]dt\,, \\
\nonumber v_{\ell}(t)&=&\frac{1}{2\pi}\int
V_{\ell}(\omega)\exp[i\omega t]d\omega\ .
\end{eqnarray}
In the zeroth order (linear) approximation
$({\mathcal{N}}_{\omega}\rightarrow 0) $
\begin{equation}
V_{\ell}^{(0)}(\omega)=\frac{\gamma_p U_{\ell}(\omega)}
{i\omega+\gamma_p}\,,\label{52}
\end{equation}
which is the simplified version of Eq.~(\ref{eq:v-u-lin-ko}). This
allows us to find in the linear approximation
\begin{equation}
\langle v_{\ell}^{2}(t)\rangle =\int\frac{d\omega}{2\pi}
{\mathcal{F}}_{\ell} (\omega)=\int\frac{d\omega}{2\pi}
\frac{\gamma_p^{2}\overline{F_{u,\ell}(\omega)}}
{\omega^{2}+\gamma_p^{2}}\, , \label{53}
\end{equation}
where $F_{u,\ell}(\omega)$ is the correlation functions of
$U_{\ell}(\omega) $:
\begin{equation}
2\pi\delta(\omega+{\omega}^{\prime})F_{u,\ell} (\omega)=\langle
U_{\ell} (\omega)U_{\ell}({\omega}^{\prime})\rangle\,, \label{54}
\end{equation}
similarly to Eq.~(\ref{eq:def-Fu}).

In the limit $\tau_p\gg \ell/u_* $ one can neglect in
Eq.~(\ref{53}) the $\omega $-dependence of
$\overline{F_{u,\ell}(\omega)} $, which has characteristic width
$\ell/u_* $ and conclude:
\begin{eqnarray} \nonumber
v^{2}_{\ell,0}&\equiv&\langle v_{\ell}^{2}(t)\rangle
\approx\frac{\gamma_p}{2} \overline{F_{u,\ell}(0)}\approx
u^{2}_{\ell} \frac{\ell\, \gamma_p} {u_*}\approx u^{2}_{\ell}
\frac\ell{\ell_*}, \\ u^{2}_{\ell}&\equiv&\langle
u_{\ell}^{2}(t)\rangle\,, \label{55}
\end{eqnarray}
in agreement with Eq.~(\ref{eq:v-u-1}).

\paragraph{First nonlinear correction.}
To evaluate the first nonlinear correction to
Eq.~(\ref{55}) one has to substitute $V_{\ell} (\omega) $ from
Eq.~(\ref{52}) into Eq.~(\ref{eq:nonlin}) for
${\mathcal{N}}_{\omega}$:
\begin{eqnarray}\nonumber
&&V_{\ell,1}(\omega)= -\frac{\gamma^{2}_p}{2\pi \ell}
\int d\omega_1d\omega_{2}\delta(\omega+\omega_1+\omega_{2})\\
&\times& \frac{U_{\ell}
(\omega_1)}{i\omega_1+\gamma_p}\frac{U_{\ell}(\omega_{2})}
{i\omega_{2}+\gamma_p}\ .\label{56}
\end{eqnarray}
Using Eq.~(\ref{56}) instead of Eq.~(\ref{52}) we obtain instead
of Eq.~(\ref{53})
\begin{eqnarray}\label{57}
v^{2}_{\ell,1}&\equiv& \langle [v_{\ell,1}(t)]^2 \rangle =
\int\frac{d\omega}{2\pi}{F_{u,\ell ,1}} (\omega)\, ,\\
F_{u,\ell ,1} (\omega)& = & \frac{2\gamma^{4}_p}
{(\omega^{2}+\gamma_p^{2})\ell^{2}}
\int\frac{d\omega_1d\omega_{2}}{2\pi}
\nonumber \\
&& \times \frac{\overline{F_{u,\ell} (\omega_1)} \
\overline{F_{u,\ell} (\omega_{2})}\delta(\omega+\omega_1
+\omega_{2})}{(\omega^{2}_1+\gamma_p^{2})
(\omega^{2}_{2}+\gamma_p^{2})}\ . \nonumber
\end{eqnarray}
In this derivation we assumed for simplicity the Gaussian
statistics of the velocity field.

Now let us estimate
\begin{equation}
v^{2}_{\ell,1} \approx\frac{\left[\overline{F_{u,\ell}(0)}\right
]^{2}}{\ell^{2}}\approx \frac {u^{4}_{\ell}}{u_*^2 }\approx
u^{2}_{\ell}\left( \frac{\ell}{\ell_*}\right)^{2/3}\, , \label{58}
\end{equation}
that is much larger than the result (\ref{55}) for $v^{2}_{\ell,0} $
obtained in the linear approximation. This means that the simple
iteration procedure we used is inconsistent, since it involves
expansion in large parameter [$(\ell_*/\ell)^{1/3}$].

\paragraph{Renormalized perturbative expansion.}
A similar situation with a perturbative expansion occurs in the
theory of hydrodynamic turbulence, where a simple iteration of the
nonlinear term with respect to the linear (viscous) term, yields
the power series expansion in ${\mathcal{R}}e^{2}\gg 1 $. A way
out, used in the theory of hydrodynamic turbulence is the
Dyson-Wyld re-summation of one-eddy irreducible diagrams (for
details see, e.g., Refs.~\cite{LP0,LP1,LP2}). This procedure
corresponds to accounting for the nonlinear (so-called "turbulent"
viscosity) instead of the linear, kinematic viscosity. A similar
approach in our problem implies that we have to account for the
self-consistent, nonlinear renormalization of the particle
frequency $\gamma_p\Rightarrow\Gamma_p(\ell) $ in
Eq.~(\ref{eq:nonlin}) and to subtract the corresponding terms from
$\tilde{{\mathcal{N}}}_{\omega} $. With these corrections,
Eq.~(\ref{eq:nonlin}) reads:
\begin{eqnarray}
[\,i\omega+\Gamma_p(\ell)\,]V_{\ell}(\omega)=\gamma_p U_{\ell}
(\omega)+\tilde{{\mathcal{N}}}_{\omega}\ . \label{eq:Ncor}
\end{eqnarray}
Here $\tilde{{\mathcal{N}}}_{\omega} $ is the nonlinear term
${\mathcal{N}}_{\omega} $ after substraction of the nonlinear
contribution to the difference
\begin{equation}
\Delta_p\equiv\Gamma_p(\ell)-\gamma_p
\approx\frac{v_{\ell}^{2}/\ell^{2}}{\Gamma_p(\ell)} \ . \label{60}
\end{equation}
The latter relation actually follows from a more detailed
perturbation diagrammatic approach. In our context it is
sufficient to realize that in the limit $\Gamma_p(\ell)\gg\gamma_p
$ one may evaluate $\Gamma_p(\ell) $ by a simple dimensional
reasoning:
\begin{equation}
\Gamma_p(\ell)\approx v_{\ell}/\ell\, ,\label{61}
\end{equation}
which is consistent with Eq.~(\ref{60}). In addition,
Eq.~(\ref{60}) has a natural limiting case
$\Gamma_p(\ell)\rightarrow\gamma_p $ when
$v_{\ell}/\ell\ll\gamma_p $. Now using Eq.~(\ref{eq:Ncor}) instead
of Eq.~(\ref{52}) we arrive at:
\begin{equation}
V_{\ell}^{(0)}(\omega)=\frac{\gamma_p U_{\ell}
(\omega)}{i\omega+\Gamma_p(\ell)}\ .\label{62}
\end{equation}
Accordingly, instead of the estimates~(\ref{55}) one has:
\begin{equation}
\tilde{v}^{2}_{\ell,0}\approx u^{2}_{\ell}\,\frac{\gamma_p^{2}\,
\ell} {\Gamma_p(\ell)\,u_* }\approx \,u^{2}_{\ell}\,
\frac{\gamma_p}{\Gamma_p(\ell)} \left(\frac{\ell}{\ell_*}\right)\,
. \label{63}
\end{equation}
The latter equation together with Eq.~(\ref{61}) allows to
evaluate $\Gamma_p(\ell)$ as follows:
\begin{equation}
\Gamma_p(\ell)\approx\left (\frac{\gamma_p^{2} u_\ell^2}{ \ell\,
u_* }\right )^{1/3}\approx
\gamma_p\left(\frac{\ell_*}{\ell}\right)^{1/9} \ .\label{64}
\end{equation}
Hence the estimate~(\ref{63}) becomes
\begin{equation}
\tilde{v}^{2}_{\ell,0}\approx u^{2}_{\ell} \left(\frac{\ell}
{\ell_*}\right)^{10/9} \approx u^{2}_{\ell}
\left[\frac{\tau(\ell)} {\tau_p}\right]^{5/3} \ .\label{65}
\end{equation}
Repeating the evaluation of the nonlinear correction
$\tilde{v}^{2}_{\ell,2} $ with the renormalized Eq.~(\ref{eq:Ncor}) we
find that
\begin{equation}
\tilde{v}^{2}_{\ell,1}\approx\tilde{v}^{2}_{\ell,0}\ . \label{66}
\end{equation}
This means that now the expansion parameter is of the order of 1,
in accordance with  the renormalized perturbation approach.

\section{The clustering instability of the second moment
of particle number density} \label{moments}

In this section we will present a quantitative analysis for the
clustering instability of the second moment of particles number
density.

\subsection{Basic equations}
\label{bas}

To determine the growth rate of the clustering instability let us
consider the equation for the two-point correlation function
$\Phi(t, \B R)$ of particle number density:
\begin{eqnarray}
{\partial \Phi \over \partial t} = [B(\B R) + 2 {\B U}(\B R)\cdot
\B {\nabla} + \hat D_{\alpha \beta}(\B R)
\nabla_{\alpha} \nabla_{\beta}] \, \Phi(t,\B R) \,, \nonumber\\
\label{WW6}
\end{eqnarray}
(see Ref.~\cite{02-Our-Clust-PRE}). The meaning of the
coefficients $ B(\B R) $, $\B U(\B R) $ and $ \hat D_{\alpha
\beta}(\B R)$ is as follows:

The function $ B(\B R) $ is determined by the compressibility of
the velocity field and it causes the generation of fluctuations of
the number density of particles.

The vector $ \B U(\B R) $ determines a scale-dependent drift
velocity which describes a transport of fluctuations of particle
number density from smaller scales to larger scales, i.e., in the
regions with larger turbulent diffusion. The latter can decrease
the growth rate of the clustering instability. Note that $ {\bf
U}(\B R=0) = 0 $ whereas $ B(\B R=0) \not= 0 .$ For incompressible
velocity field $ {\bf U}(\B R) = 0 $ and $ B(\B R) = 0$.

The scale-dependent tensor of turbulent diffusion $
\hat{D}_{\alpha\beta}(\B R)$ is also affected by the
compressibility. In very small scales this tensor is equal to the
tensor of the molecular (Brownian) diffusion, while in the
vicinity of the maximum scale of turbulent motions this tensor
coincides with the regular tensor of turbulent diffusion.

Thus, the clustering instability is determined by the competition
between these three processes. The tensor
$\hat{D}_{\alpha\beta}(\B R)$ may be written as
\begin{eqnarray}
\hat{D}_{\alpha\beta}(\B R) &=& 2 D \delta_{\alpha\beta} + D \Sp
T_{\alpha \beta}(\B R)\,, \br D_{\alpha \beta}\Sp T (\B R) &=&
\tilde{D}_{\alpha \beta}\Sp T (0) - \tilde{D} _{\alpha \beta}\Sp T
(\B R)\ .
\end{eqnarray}
The form of the coefficients $ B(\B R) $, $\B U(\B R) $ and $ \hat
D_{\alpha \beta}(\B R)$ depends on the model of turbulent velocity
field. For instance, for the random velocity with Gaussian
statistics of the Lagrangian trajectories $\B \xi(\B r|t)$ these
coefficients are given by
\begin{eqnarray}
\label{LL30} B(\B R) &\approx& 2 \int_{0}^{\infty} \langle b[0,\B
\xi(\B r_1|0)] b[\tau,\B \xi(\B r_2|\tau)] \rangle \,d \tau\,, \br
{\B U}(\B R) & \approx & - 2 \int_{0}^{\infty} \langle {\B v}[0,\B
\xi(\B r_1|0)] b[\tau,\B \xi(\B r_2|\tau)] \rangle \,d\tau \,, \br
\tilde D_{\alpha \beta}\Sp T (\B R) &\approx& 2 \int_{0}^{\infty}
\langle v_{\alpha}[0,\B \xi(\B r_1|0)] v_{\beta}[\tau,\B \xi(\B
r_2|\tau)] \rangle \,d \tau \,
\end{eqnarray}
(for details see Ref.~\cite{02-Our-Clust-PRE}), where $
b=\text{div} \, {\mathbf{v}} .$

\subsection{Clustering instability}
\label{mod-a}

Let us study the clustering instability. Consider the range of
scales $a \leq \ell \ll \ell_\ast$, where the size of a particle $
a \geq \eta .$ Then the relationship between
$\tilde{v}^{2}_{\ell,0}$ and $u^{2}_{\ell}$ reads:
\begin{equation}
\tilde{v}^{2}_{\ell,0} = u^{2}_{\ell} \left[\frac{\tau(\ell)}
{\tau_p}\right]^{s} \,, \label{L1}
\end{equation}
where according to Eq.~(\ref{65}) the exponent $s=5/3 .$ In this
case the expression for the turbulent diffusion tensor in
nondimensional form reads
\begin{eqnarray} \label{W12}
&& D^{^{\rm T}}_{\alpha \beta}(\B R)= R^{(4s-7)/3} (C_{1} R^{2}
\delta_{\alpha \beta} + C_{2} R_\alpha R_\beta ) \,,
\\ \nonumber
&&C_{1} = { 5 + 4 s + 6 \sigma\Sb{T} \over 9 \, (1 +
\sigma\Sb{T})}\,,
\\ \nonumber
&& C_{2} = {(4 s - 1) (2\sigma\Sb{T} - 1) \over 9 \,( 1 +
\sigma\Sb{T})} \,,
\end{eqnarray}
where $R$ is measured in the units of $L$ and time $t$ is measured
in the units of $\tau_{_{L}} \equiv \tau(\ell=L)$. The parameter
$\sigma\Sb{T}$ is defined by analogy with Eq.~(\ref{19}):
\begin{eqnarray}
\label{S} \hskip -0.4cm \sigma\Sb{T}\equiv \frac{\B \nabla \cdot
\B D\Sb T\cdot \B \nabla }{\B \nabla \times \B D\Sb T\times \B
\nabla }= \frac{\nabla_\alpha \nabla_\beta D^{^{\rm T}}_{\alpha
\beta}(\B R)}{\nabla_\alpha\nabla_\beta D^{^{\rm T}} _{\alpha'
\beta'}(\B R) \epsilon_{\alpha\alpha'\gamma}
\epsilon_{\beta\beta'\gamma} }\,,
\end{eqnarray}
where $ \epsilon_{\alpha\beta\gamma}$ is the fully antisymmetric
unit tensor. Equations~(\ref{19}) and~(\ref{S}) imply that $
\sigma\Sb{T}=\sigma_v$ in the case of $\delta$-correlated in time
compressible velocity field.

For a random incompressible velocity field with a finite
correlation time the tensor of turbulent diffusion $ \tilde
D^{^{\rm T}}_{\alpha \beta} (\B R) = \tau^{-1} \langle
\xi_\alpha(\B r_1|t) \xi_\beta(\B r_2|t) \rangle $ (see
Ref.~\cite{02-Our-Clust-PRE}) and the degree of compressibility of
this tensor is
\begin{eqnarray}
\label{SN} \sigma\Sb{T} = {\langle (\B \nabla \cdot \B {\xi})^{2}
\rangle \over \langle (\B \nabla {\bf \times} \B {\xi})^{2}
\rangle} \,,
\end{eqnarray}
where $\B \xi(\B r|t)$ is the Lagrangian trajectory.

To determine the functions $B(\B R)$ and $\B U(\B R)$ we use the
general form of the two-point correlation function of the particle
velocity field in the the range of scales $\eta \leq \ell \ll
\ell_\ast$:
\begin{eqnarray}
\label{L3} \langle v_{\alpha}(t,\B r) v_{\beta}(t&+&\tau,\B r + \B
R) \rangle = {1 \over 3} [\delta_{\alpha \beta} - (C_{1}^v R^{2}
\delta_{\alpha \beta}
\nonumber\\
&& + \, C_{2}^v R_\alpha R_\beta)\, R^{2(s-2)/3}] f(\tau) \,,
\\ \nonumber
&& C_{1}^v = { (4 + s + 3 \sigma_v)\over 3 \, (1 + \sigma_v)}\,,
\\ \nonumber
&& C_{2}^v = {(1 + s) (2\sigma_v - 1)    \over 3 \,( 1 +
\sigma_v)} \, .
\end{eqnarray}
Substitution Eq.~(\ref{L3}) into Eq.~(\ref{LL30}) yields
\begin{eqnarray}
\label{L2} \B U(\B R) = U_0 \, R^{(4s-7)/3} \,,
\\\nonumber
B(\B R) = B_0 \, R^{(4s-7)/3} \,,
\end{eqnarray}
where
\begin{eqnarray*}
U_0 = \beta \, {\sigma_v \over \sigma_v + 1} \,, \quad B_0 =
\alpha \, U_0 \,
\end{eqnarray*}
and the coefficients $\alpha$ and $\beta$ depend on the properties
of turbulent velocity field.  The dimensionless functions   $B_0$
and $U_0$  in Eq.~(\ref{L2}) are measured in the units of
$\tau_{_{L}}^{-1}$.

For the $ \delta $-correlated in time random Gaussian compressible
velocity field $\sigma\Sb{T}=\sigma_v$ and
\begin{eqnarray}
&& B(\B R) = \nabla_{\alpha} \nabla_{\beta} \hat D_{\alpha
\beta}(\B R) \,, \label{WL6} \\
&& U_{\alpha}(\B R) = \nabla_{\beta} \hat D_{\alpha \beta}(\B R)
\label{WLL6}
\end{eqnarray}
(for details see Ref.~\cite{02-Our-Clust-PRE,15,19}). In this case
the second moment $ \Phi(t,\B R) $ can only decay, in spite of the
compressibility of the velocity field.

For the finite correlation time of the turbulent velocity field
$\sigma\Sb{T} \not=\sigma_v$ and Eqs.~(\ref{WL6}) and~(\ref{WLL6})
are not valid. The clustering instability depends on the ratio
$\sigma\Sb{T} / \sigma_v$. In order to provide the correct
asymptotic behaviour of Eq.~(\ref{L2}) in the limiting case of the
$ \delta $-correlated in time random Gaussian compressible
velocity field we have to choose the coefficients $\beta$ and
$\alpha$ in the form:
\begin{eqnarray*}
\beta = 8 (4 s^2 + 7 s - 2) / 27 \,, \, \, \alpha = (4 s + 2 ) / 3
\, .
\end{eqnarray*}
Note that when $s< 1/4$ the parameters $\beta < 0 $ and $B(\B R) <
0$. In this case there is no clustering instability of the second
moment of particle number density.

Thus, Eq.~(\ref{WW6}) in a non-dimensional form reads:
\begin{eqnarray}
{\partial \Phi \over \partial t} &=& R^{(4s-7)/3} [R^2 \Phi''
(C_{1} + C_{2}) + 2\, R \Phi' (U_0 + C_{1})
\nonumber\\
&&+ B_0 \Phi] \, . \label{W14}
\end{eqnarray}
Consider a solution of Eq.~(\ref{W14}) in the vicinity of the
thresholds of the excitation of the clustering instability, where
$(\partial \Phi / \partial t) R^{(7-4s)/3}$ is very small. Thus,
the solution of~(\ref{W14}) in this region is
\begin{eqnarray}\label{L6}
\Phi(R) = A_{1} R^{-\lambda_1} \,,
\end{eqnarray}
where $ \lambda_1 = \lambda \pm i \mu ,$
\begin{eqnarray}\nn
\lambda &=& {C_{1} - C_{2} + 2 U_0 \over 2 (C_{1} + C_{2})} \,,
\quad \mu = {C_3 \over 2 (C_{1} + C_{2})} \,, \br C_3^2&=&4
B_0(C_1+C_2) - (C_{1} - C_{2} + 2 U_0)^{2} \ .
\end{eqnarray}
Since the correlation function $ \Phi(R) $ has a global maximum at
$ R = a ,$ the coefficient $ \, C_{1} > C_{2}- 2U_0$ if $\mu$ is a
real number (see below). Thus the asymptotic solution of the
equation for the two-point correlation function $\Phi(t, \B R)$ of
the particle number density in the range of scales $a \leq \ell
\leq \ell_\ast$ reads
\begin{eqnarray}\label{L8}
\Phi(R) = A_{1} R^{-\lambda} \sin[ \mu \ln (\ell_\ast / R)] \,,
\end{eqnarray}
where
\begin{eqnarray*}
A_{1} = \biggl({L \over a}\biggr)^\lambda {1 \over \sin[ \mu \ln
(\ell_\ast / a)]} \, .
\end{eqnarray*}

Now consider the range of scales $\ell_\ast \ll \ell \ll L$. In
this case the nondimensional form of the turbulent diffusion
tensor is given by
\begin{eqnarray} \label{L4}
&& D^{^{\rm T}}_{\alpha \beta}(\B R)= R^{-2 / 3} (\tilde C_{1}
R^{2} \delta_{\alpha \beta} + \tilde C_{2} R_\alpha R_\beta) \,,
\\ \nonumber
&&\tilde C_{1} = { 2 (5 + 3 \tilde \sigma\Sb{T})\over 9 \, (1 +
\tilde \sigma\Sb{T})}\,, \; \; \tilde C_{2} = {4 (2 \tilde
\sigma\Sb{T} - 1) \over 9 \,( 1 + \tilde \sigma\Sb{T})} \,,
\end{eqnarray}
and Eq.~(\ref{WW6}) reads:
\begin{eqnarray}
{\partial \Phi \over \partial t} &=& R^{-2 / 3} [R^2 \Phi''
(\tilde C_{1} + \tilde C_{2}) + 2\, R \Phi' \tilde C_{1}] \, .
\label{L5}
\end{eqnarray}
Here we took into account that in this range of scales the
functions $ B(\B R) $ and $\B U(\B R) $ are negligibly small, and
the degree of compressibility of the turbulent diffusion tensor
can be different in the above two ranges of scales.

Consider a solution of Eq.~(\ref{L5}) in the vicinity of the
thresholds of the excitation of the clustering instability, when
$(\partial \Phi / \partial t) R^{2/3}$ is very small. Thus, the
solution of~(\ref{L5}) is given by
\begin{eqnarray}\label{L7}
\Phi(R) = A_{2} R^{-\lambda_2} \,,
\end{eqnarray}
where
\begin{eqnarray}\nn
\lambda_2 = {|\tilde C_{1} - \tilde C_{2}| \over \tilde C_{1} +
\tilde C_{2}} = {|7 - \tilde \sigma\Sb{T}| \over 3 + 7 \tilde
\sigma\Sb{T}} \, .
\end{eqnarray}

The growth rate of the second moment of particle number density
can be obtained by matching the correlation function $ \Phi(R) $
and its first derivative $ \Phi'(R) $ at the boundary of the above
two ranges of scales, i.e., at the points $R = \ell_\ast / L .$
The matching yields that in the range of scales $\ell_\ast \ll
\ell \leq L$ the asymptotic solution of the  equation for the
two-point correlation function $\Phi(t, \B R)$ has the form of
Eq.~(\ref{L7}) with
\begin{eqnarray*}
A_{2} = (-1)^k A_{1} \biggl({\ell_\ast \over L}\biggr)^{\lambda_2
- \lambda + 1} {\mu \over \lambda_2} \, .
\end{eqnarray*}

Such matching is possible only when $ \lambda_1 $ is a complex
number, i.e., when $ C_3^2 > 0$ (i.e., $\mu$ is a real number).
The latter determines the necessary condition for the clustering
instability of particle spatial distribution. The range of
parameters $ (\sigma_v , \sigma\Sb{T})$ for which the clustering
instability of the second moment of particle number density may
occur is shown in Fig. 1. The line $ \sigma_v = \sigma\Sb{T} $
corresponds to the $ \delta $-correlated in time random
compressible velocity field for which the clustering instability
cannot be excited. The various curves indicate results for
different value of the parameter $s$. In particular, the value $s
= 7/4$ corresponds to the turbulent diffusion tensor with the
scaling $\propto R^2$ [see Eq.~(\ref{W12})]. The curves for $s =
7/4$ (dashed) and $s=5/3$ (solid) practically coincide. The
parameter $s$ can be considered as a phenomenological parameter,
and the change of this parameter from $s=5/3$ to $s=0$ can
describe a transition from one asymptotic behaviour (in the range
of scales $a \leq \ell \leq \ell_\ast$) to the other ($\ell_\ast
\ll \ell \leq L$).

\begin{figure}
\centering
\includegraphics[width=8cm]{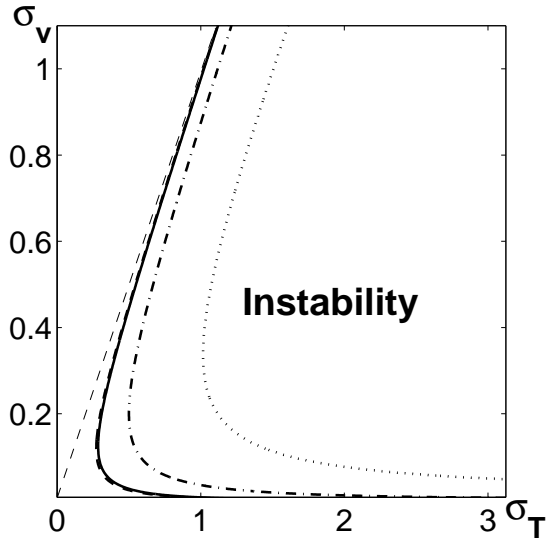}
\caption{\label{Fig1} The range of parameters $ (\sigma_v ,
\sigma\Sb{T})$ for which the clustering instability may occur. The
various curves indicate results for $s = 7/4$ (dashed), $s=5/3$
(solid), for $s=1$ (dashed-dotted) and for $s=2/3$ (dotted). The
thin dashed line $ \sigma_v = \sigma\Sb{T} $ corresponds to the $
\delta $-correlated in time random compressible velocity field.}
\end{figure}

\section{Discussion}
\label{s:disc}

Formation and evolution of particle clusters are of fundamental
significance in many areas of environmental sciences, physics of
the atmosphere and meteorology (e.g., smog and fog formation, rain
formation), see e.g., Ref.~\cite{S86,FS88,PK97,VY00}, transport
and mixing in industrial turbulent flows (like spray drying and
cyclone dust separation, dynamics of fuel droplets), see e.g.,
Ref.~\cite{98CST,A1,A2,A3,00Bel,23} and references therein. The
analysis of the experimental data showed that the spatial
distributions of droplets in clouds are strongly inhomogeneous
\cite{4}. The small-scale inhomogeneities in particle distribution
were observed also in laboratory turbulent flows \cite{5,7}.

The analyzed effect of particle clustering may be of relevance in
turbulent fluid flows of different nature with inertial particles
or droplets (e.g., in atmospheric turbulence, combustion and in a
laboratory turbulence). In particular, this effect can cause
formation of small-scale inhomogeneities in spatial distribution
of fuel droplets in diesel engines. The starting point is the
above theoretical model that describes the inertial particle
clustering. For numerical estimates we adopt the operating
parameters for a typical diesel engines, taking the crankshaft
rotational speed $N $=3000 rev/min, that corresponds to the
average piston speed, ${\bar{v}}_{p}=10^{3} $ cm/s, and the
turbulent velocity ${v}^{\prime}\simeq5\cdot10^{2} $ cm/s. Taking
into account that the spatial integral space (the characteristic
scale of the largest turbulent eddies) for this conditions is
$L\simeq\,0.3 \,$ cm (see Ref.~\cite{A3}), we find for the
Reynolds number:
\begin{eqnarray}\label{eq:Re}
{\mathcal{R}}e&=&\frac{{v}^{\prime}L}{\nu}\simeq 10^4\,, \quad
\text{with}\\ \nonumber {v}^{\prime}&\simeq& 5\cdot10^{2}
\text{cm/sec.},\quad L\simeq\,0.3 \, \text{cm}\ .\label{73}
\end{eqnarray}
Here the kinematic viscosity of fluid is $\nu\approx0.015 \,
$cm$^{2}$/s. To estimate $\nu$ we used the following values for
the fluid density $\rho=0.025\, $g/cm$^{3} $ (at 62 atm. and
$T$=860 K), and for kinematic viscosity
\begin{equation}
\nu=\nu_{0}R^{(3-\gamma)/2},\label{74}
\end{equation}
where $R=(\rho/\rho_{0})\approx 15 $ is the compression ratio.
Then the Kolmogorov micro-scale is $\eta\approx
{L}/{\mathcal{R}}e^{3/4}\sim 3 \, \mu$m. A more accurate estimate
is
\begin{equation}
\eta\approx\frac{L}{({\mathcal{R}}e/{\mathcal{R}}e_\text{cr})^{3/4}}
\,,\label{75}
\end{equation}
where ${\mathcal{R}}e_\text{cr}>1$ is the critical value of
${\mathcal{R}}e$ at which the laminar flow becomes unstable.
Taking ${\mathcal{R}}e_\text{cr}\approx 5 $, we obtain a more
realistic value $\eta \simeq 10 \,\mu$m.

Note that turbulence in the in diesel engines is neither
homogeneous nor isotropic. However, we study dynamics of droplets
at very small scales which are of the order of tens of hundreds
microns. In such small scales the turbulence can be considered as
a quasi isotropic and homogeneous. Anisotropy and inhomogeneities
of turbulence can be essential in scales which are of the order of
1 cm and larger.

Consider a region of unstable cluster scales: $\eta < \ell <
\ell_* $. Assuming the fuel density in droplets
$\rho_\text{dr}=0.85 \, \text{g/cm} ^{3} $, we find from the
expression (\ref{47}) the estimate for the critical value of the
droplet radius, $a_{\ast}\approx 4\, \mu $m, indicating that all
droplets with $a> 4\, \mu $m are unstable against formation of the
clusters with scales ranging from $\eta$ to $\ell_*$. This
equation yields the following values of $\ell_*$: $\ell_* \approx
160\,\mu $m for $a=10 \, \mu $m and $\ell_* \approx 1$cm for
$a=40\, \mu $m.

The characteristic time of the cluster growth $\tau_{cl}({ \ell})
= 1/\gamma _{cl}$ depends on the cluster size $\ell$. Estimate for
$\ell=\ell_*/3$ gives $\tau_\text{cl}(\ell_*/3)\approx 1.7
\tau(\ell_*)\approx \tau_\text{dr}$. The turnover time
$\tau(\ell)$ can be estimated as follows:
\begin{equation}
\tau(\ell)\simeq 5\left(\frac{L}{v^{\prime}}
\right)\left(\frac{\ell}{L}\right)^{2/3}\simeq 3\times
10^{-3}\left(\frac{\ell}{L}\right)^{2/3} {\rm sec} \ . \label{77}
\end{equation}
Here we used our estimates~(\ref{eq:Re}) for the integral scale
$L$ and turbulent velocity $v^\prime$. Assuming for the droplets
with the size $a=10 \, \mu$m, $\ell=\ell_*/3\approx 50 \, \mu$m
(with $\ell_*=160 \, \mu$m) one has $\tau_\text{cl}(\ell)\approx
0.35 \, \mu$s. For the droplets with the radius $a=40\,\mu$m with
$\ell_*\approx 1$ cm, one has $\tau_\text{cl}({ \ell_*}/3)\approx
5.5 \,  \mu$s, which is larger than the characteristic spray time
in the combustion chamber (a few milliseconds). These estimates
imply that for the chosen parameters the distribution of the
droplets in the spray will be shifted to the droplets with larger
diameters at the given turbulent intensity.

Now we estimate the time of the turbulent air-fuel mixing,
$\tau_{_\text{T}} \simeq 1/\gamma_{_\text{T}} $ in the combustion
chamber. Taking into account that
\begin{equation}
\gamma_{_\text{T}}
=\frac{D_{_\text{T}}}{d^{2}}=\frac{v^{\prime}L}{3d^{2}}\,,\label{78}
\end{equation}
where $D_{_\text{T}} $ is the coefficient of turbulent diffusion,
and $d\simeq10 \,$ cm is the bore diameter, we find for the time
of turbulent air-fuel mixing $\tau_{_\text{T}} \approx 1\,$s.
Thus, the turbulent diffusion itself is too slow and a strong
external global flow - "swirl" is needed for the effective
air-fuel mixing.

In the previous sections we have shown that the particle spatial
distribution in the turbulent flow field is unstable against
formation of clusters with particle number density that is much
higher than the average particle number density. Obviously this
exponential growth at the linear stage of instability should be
saturated by nonlinear effects. The nonlinear saturation for the
water droplets in the atmospheric turbulence (in clouds) was
discussed in Ref.~\cite{02-Our-Clust-PRE}. Here we discuss a
similar issue for the fuel droplets under the typical conditions
pertinent to diesel engines.

A momentum coupling of particles and turbulent fluid is essential
when $m_p n_\text{cl}\sim\rho $, i.e., the mass loading parameter,
$\phi=m_p n_\text{cl}/\rho$, is of the order of unity (see, e.g.,
Ref.~\cite{LOP}). This condition implies that the kinetic energy
of air $\rho \langle{\mathbf{u}}^{2}\rangle $ is of the order of
the droplets kinetic energy $m_p n_ \text{cl} \langle
{\mathbf{v}}^{2}\rangle $, where ${|{\mathbf{u}}|
\sim|{\mathbf{v}}|} $. This yields:
\begin{equation}
n_\text{cl}\sim a^{-3}(\rho/3\rho_p)\ . \label{71}
\end{equation}
For the fuel droplets in the diesel combustion chambers (see, e.g.
Ref.~\cite{A1}) $\rho_p/\rho \simeq 34 $. Thus, e.g., for $a
\simeq 10 \, \mu$m we obtain $n_\text{cl}\sim 3\times 10^7$
cm$^{-3}$. For the cluster with the size $\ell\approx
\ell_*/3\approx 50 \, \mu$m this yields for the total number of
particles in the cluster of that size $N_\text {cl}\simeq
\ell^{3}n_\text{cl}\sim 30 $. Note that the mean number density of
droplets in a combustion camera $\bar{n} $ is about $10^4 $
cm$^{-3} .$ Therefore the clustering instability of droplets in
the diesel engines increases their concentrations in the clusters
by the orders of magnitude.

Note that in the initial stage of the clustering instability, when
the mass loading parameter is small, i.e., for the small number
density of droplets, we can neglect the droplets collisions, and
consider only the collisions between the droplets and the
molecules of the surrounding fluid (air). In the nonlinear stage
of the clustering instability, the four-way coupling can be
effective, and the kinetic approach for analysis of the droplets
collisions inside the cluster can be important. Moreover, the
coagulations of droplets due to their collisions change the size
of the droplets. This effect is determined by the kinetic
Smoluchovsky equation for droplets size distribution. However, the
interaction between droplets and the surrounding fluid even at
this stage of the clustering instability can be described on the
level of the continuum hydrodynamic approach. The main purpose of
the present paper is to describe the initial (linear) stage of the
clustering instability and to determine the conditions for the
onset of the clustering instability of fuel droplets. This stage
of the instability can be analyzed only using the continuum
hydrodynamic approach because the mean free path $l_c$ of the
molecules of the surrounding fluid is much smaller than all the
characteristic spatial scales in the problem (e.g., the Kolmogorov
micro-scale $\eta$, the size of droplets $a$, etc). Indeed, for
diesel engines these parameters are: $l_c = 0.01 \, \mu$m, $\eta=
3 \, \mu$m, $a \geq 10 \, \mu$m. Similarly, the mean time $\tau_c$
between collisions of the molecules of the surrounding fluid is
much smaller than all the characteristic time scales (e.g., the
correlation time in the Kolmogorov micro-scale $\tau_\eta$, the
Stokes time for droplets $\tau_p$, etc). Indeed, for diesel
engines these parameters are $\tau_c = 2 \times 10^{-11} \, $s,
$\tau_\eta= 3 \times 10^{-5} \, $s, and $\tau_p = 10^{-4} \, $s
for $a = 10 \, \mu$m. At the strongly nonlinear stage of the
clustering instability, when the mean free path of the droplets
inside a cluster becomes much smaller than the size of the
cluster, the continuum hydrodynamic approach can be still used to
obtain a rough estimate of the number density of droplets inside
the cluster at the saturation of the clustering instability.

\begin{acknowledgements}
This research was supported in part by The German-Israeli Project
Cooperation (DIP) administrated by the Federal Ministry of
Education and Research (BMBF), by the Israel Science Foundation
governed by the Israeli Academy of Science, by Swedish Ministry of
Industry (Energimyndigheten, contract P 12503-1), by the Swedish
Royal Academy of Sciences, the STINT Fellowship program.
\end{acknowledgements}


\begin{references}

\bibitem{10}  A.S. Monin and  A.M. Yaglom, \textit{Statistical Fluid Mechanics:
Mechanics of Turbulence} (M.I.T. Press, Cambridge, 1975), vol. 2.

\bibitem {C80}
G.T. Csanady, {\em Turbulent Diffusion in the Environment}
(Reidel, Dordrecht, 1980).

\bibitem {PS83}
F. Pasquill and F.B. Smith, {\em Atmospheric Diffusion} (Ellis
Horwood, Chichester, 1983).

\bibitem {McC90} W.D. McComb, {\em The Physics of Fluid
Turbulence} (Clarendon Press, Oxford, 1990).

\bibitem {S96} D. Stock, {\it J. Fluids Engineering} \textbf{118}, 4 (1996).

\bibitem {BL97} A.K. Blackadar, {\em Turbulence and Diffusion
in the Atmosphere} (Springer, Berlin, 1997).

\bibitem {W00} Z. Warhaft, Annu. Rev. Fluid Mech. {\bf 32}, 203
(2000).

\bibitem {S01} B. Sawford, Annu. Rev. Fluid Mech. {\bf 33}, 289
(2001).

\bibitem {BH03} R.E. Britter and S.R. Hanna, Annu. Rev. Fluid
Mech. {\bf 35}, 469 (2003).

\bibitem{5} J.R. Fessler, J.D. Kulick and J.K. Eaton,
Phys. Fluids \textbf{6}, 3742 (1994).

\bibitem{7} A. Aliseda, A. Cartellier, F. Hainaux and  J.C. Lasheras,
J. Fluid Mech. \textbf{468}, 77 (2002).

\bibitem{4} R.A. Shaw, Ann. Rev. Fluid Mech. \textbf{35}, 183
(2003).

\bibitem{96-PRL-EKR} T. Elperin, N. Kleeorin  and I. Rogachevskii,
Phys. Rev. Lett. \textbf{77}, 5373 (1996).

\bibitem{02-Our-Clust-PRE} T. Elperin, N. Kleeorin, V. L'vov,
I. Rogachevskii and D. Sokoloff, Phys. Rev. E \textbf{66}, 036302
(2002). Also e-prints: nlin.CD/0202048 and nlin.CD/0204022.

\bibitem{14} T. Elperin, N. Kleeorin  and I. Rogachevskii,
Phys. Rev. Lett. \textbf{76}, 224 (1996).

\bibitem{14A} T. Elperin, N. Kleeorin  and I. Rogachevskii,
Phys. Rev. Lett. \textbf{81}, 2898 (1998).

\bibitem{15} T. Elperin, N. Kleeorin, I. Rogachevskii and D.
Sokoloff, Phys. Chem. Earth, \textbf{A 25}, 797 (2000).

\bibitem{19} T. Elperin, N. Kleeorin, I. Rogachevskii and D.
Sokoloff, Phys. Rev. E \textbf{63}, 046305 (2001).

\bibitem{21} R.H. Kraichnan, Phys. Fluids \textbf{11}, 945 (1968).

\bibitem {MR83}
M.R. Maxey and J.J. Riley, Phys. Fluids {\bf 26}, 883 (1983).

\bibitem {M87}
M.R. Maxey, J. Fluid Mech. {\bf 174}, 441 (1987).

\bibitem {M96}
M.R. Maxey, E.J. Chang, and L.-P. Wang, Experim. Thermal and Fluid
Science, {\bf 12}, 417 (1996).

\bibitem{9}  L.D. Landau and  E.M. Lifshits, \textit{Fluid mechanics}
(Pergamon, Oxford, 1987).

\bibitem{11} U. Frisch, \textit{Turbulence: The Legasy of A. N.
Kolmogorov} (Cambridge University Press, Cambridge 1995).

\bibitem{6} J.L. Lumley, in {\it Two-Phase and Non-Newtonian Flows,
Turbulence}, ed. P. Bradshaw, (Springer Verlag, Berlin, 1976).

\bibitem{LP0} V.S. L'vov and I. Procaccia, in \textit{Lecture Notes of
the Les Houches Summer School ``Fluctuating Geometries in
Statistical Mechanics and Field Theory"}, ed. by F. David and P.
Ginsparg, (North-Holland, Amsterdam, 1995).

\bibitem{LP1} V.S. L'vov and I. Procaccia, Phys. Rev. E
\textbf{52}, 3840 (1995).

\bibitem{LP2} V.S. L'vov and I. Procaccia, Phys. Rev. E \textbf{52},
3858 (1995).

\bibitem {S86}
J.H. Seinfeld, {\it Atmospheric Chemistry and Physics of Air
Pollution}, (John Wiley, New York, 1986), and references therein.

\bibitem{FS88}
R. Flagan and  J.H. Seinfeld, {\it Fundamentals of air pollution
engineering}, (Prentice Hall, Englewood Cliffs, 1988).

\bibitem{PK97}
H.R. Pruppacher  and J.D. Klett, {\em Microphysics of Clouds and
Precipitation}, (Kluwer Acad. Publ., Dordrecht, 1997).

\bibitem {VY00} P.A. Vaillancourt and M.K. Yau,
Bull. Americ. Meteorol. Soc. {\bf 81}, 285 (2000), and references
therein.

\bibitem{98CST} C.T. Crowe, M. Sommerfeld and Y. Tsuji,
{\it Multiphase Flows with Particles and Droplets} (CRC Press, NY,
1998).

\bibitem{A1} J.B. Heywood, {\it Internal Combustion Engine Fundamentals},
(MacGraw-Hill, Boston-New-York, 1988).

\bibitem{A2} T. Kamimoto and H. Kobayashi, Prog. Energy Combust.
Sci. \textbf{17}, 163 (1991).

\bibitem{A3} G.L. Borman and K.W. Ragland, {\it Combustion Engineering},
(MacGraw-Hill, Boston-New-York, 1999).

\bibitem{00Bel}
J. Bellan, Prog. Energy and Combust. Sci. \textbf{28}, 329 (2000).

\bibitem{23}  H.H. Chiu, Prog. Energy and Combust. Sci. \textbf{26}, 381
(2000).

\bibitem{LOP} V.S. L'vov, G. Ooms and A. Pomyalov, "Effect of particle
inertia on the turbulence in suspensions," Phys. Rev. E
\textbf{67}, 046314 (2003). Also e-print: nlin.CD/0210069.

\end{references}
\end{document}